\documentclass[12pt]{article}
\usepackage[cp1251]{inputenc}
\usepackage{amsmath}
\usepackage{amsfonts}
\usepackage{amssymb}
\def\pa{\partial}                       
\def\beq{\begin{eqnarray}}    
\def\eeq{\end{eqnarray}}      

\begin{document}

\begin{center}
{\Large\textbf{Quintic vertices  of spin 3, vector  and
scalar  fields}}

\vspace{18mm}

{\large
P.M. Lavrov$^{(a, b)} \footnote{E-mail: lavrov@tspu.edu.ru}$\;,
V.I. Mudruk$^{(c)}\footnote{E-mail:
mudruk1580@mail.ru}$}

\vspace{8mm}

\noindent  ${{}^{(a)}} ${\em
Center of Theoretical Physics, \\
Tomsk State Pedagogical University,\\
Kievskaya St.\ 60, 634061 Tomsk, Russia}

\noindent  ${{}^{(b)}} ${\em
National Research Tomsk State  University,\\
Lenin Av.\ 36, 634050 Tomsk, Russia}

\noindent  ${{}^{(c)}} ${\em
Bauman Moscow State Technical University,\\
2nd Baumanskaya St.\  5, 105005 Moscow, Russia}

\vspace{20mm}

\begin{abstract}
\noindent
As a result of special deformations of free gauge models of massless spin 3, massive vector
 and real scalar fields, quintic vertices within new approach
Buchbinder and Lavrov \cite{BL-1};
Buchbinder and Lavrov \cite{BL-2};
Lavrov \cite{L-2022} are constructed. They are described by
 local functionals which are invariant under original gauge transformations.

\end{abstract}

\end{center}

\vfill

\\

\newpage

\section{Introduction}
Beginning with famous papers of Fradkin and Vasiliev \cite{FV1,FV2} and
Vasiliev \cite{Vas1,Vas2,Vas3} construction of interactions in the theory of high spin fields
\cite{Fronsdal-1,FFrons} attracts an increasing interest due to numerous problems arising
in this
process (for recent discussions see \cite{JT,Vas,Did}).
At the present, there is strong opinion  that
the structure of cubic vertices are established very well
\cite{BBB,B,Vas4,AVas5,Mets1,BFPT,FIPT,FT,Zinov,BKTW,BIZ2}
using different methods (deformation procedure in the BV formalism
\cite{BV,BV1,BH,H}, BRST construction  \cite{BPT},
light-cone formalism \cite{Mets2}).
Studies of quartic vertices led to open new problem of locality of interactions in this order
\cite{Taron,DT,Taron1,KMP}. As to quintic vertices and vertices more of high orders
(in our knowledge) there is no some explicit results.

Recently, a reformulation of the deformation procedure in the BV formalism has  been proposed
\cite{BL-1,BL-2,L-2022}.
New method opens new possibilities in studies of interactions in the higher spin field theory
thanks to an explicit and closed form of description of deformation procedure for gauge fields.
From the point of view of the old approach based on the Noether procedure, this method seems as an
explicit summation of the infinite Taylor series in deformation parameter leading to compact
presentation of the deformed action and deformed gauge algebra.
Analysis of simple gauge systems including massless spin 3, massive vector and real scalar
fields within the new approach allowed us to study situations when in the process of
deformation the original gauge symmetry does not deform \cite{L,L1}. It was shown that cubic
vertices invariant under original gauge transformations are forbidden while
local gauge-invariant quartic vertices are explicitly found. In the present paper, we extend
the results \cite{L,L1} up to quintic vertices.

The paper is organized as follows. In section 2, quintic vertices for massless spin 3 and
real scalar fields are constructed. In section 3, quantic vertices for massless spin 3, massive
vector and real scalar fields are found.
In section 4, discussions of obtained results are given.

The DeWitt's condensed notations are used \cite{DeWitt}.
 Arguments of any functional
are enclosed in square brackets $[\;]$, and arguments of any
function are enclosed in parentheses, $(\;)$.

\section{Vertices $\sim\varphi\phi\phi\phi\phi$}
In this section, we are going to continue the research of  possible interactions
among massless spin 3 field, $\varphi^{\mu\nu\lambda}=\varphi^{\mu\nu\lambda}(x)$,
and a real scalar field, $\phi=\phi(x)$, in flat Minkowski space
of the dimension $d$  with the metric tensor $\eta_{\mu\nu}$
that was started in \cite{L}.
We begin with the initial action
\beq
\nonumber
S_0[\varphi,\phi]&=&\int dx \Big[\varphi^{\mu\nu\rho}\Box\varphi_{\mu\nu\rho}-
3\eta_{\mu\nu}\eta^{\rho\sigma}\varphi^{\mu\nu\delta}
\Box\varphi_{\rho\sigma\delta}
-3\varphi^{\mu\rho\sigma}\pa_\mu\pa^\nu\varphi_{\nu\rho\sigma}+
6\eta_{\mu\nu}\varphi^{\mu\nu\delta}\pa^\rho\pa^\sigma
\varphi_{\rho\sigma\delta}-\\
&&\qquad\quad-\frac{3}{2}\varphi_{\mu\nu\lambda}\eta^{\mu\nu}\pa^{\lambda}\pa_{\alpha}
\varphi^{\alpha\beta\gamma}\eta_{\beta\gamma}\Big]
+\int dx \frac{1}{2}\big[ \pa_{\mu}\phi\;\pa^{\mu}\phi-m^2\phi^2\big],
\label{b1}
\eeq
where the first five terms in the right hand side of (\ref{b1}) present
the Fronsdal action for massless spin $3$ field
\cite{Fronsdal-1} and
the last two terms  correspond to the action of massive scalar field.
The action (\ref{b1}) is invariant under the following gauge transformations,
\beq
\label{b2}
\delta S_0[\varphi,\phi]=0, \quad
\delta\varphi^{\mu\nu\lambda}=\pa^{(\mu}\xi^{\nu\lambda)},\quad \delta\phi=0,
\eeq
where the gauge functions $\xi^{\mu\nu}$ obey the condition $\eta_{\mu\nu}\xi^{\mu\nu}=0$.

According to general concept of \cite{BL-1,BL-2,L-2022},  the interactions
among fields are introduced with the help of anticanonical transformations
in the BV formalism
which act in the minimal antisymplectic space and transform a solution
to the classical master-equation into another. To realize the deformation procedure
for a given
dynamical system with the gauge freedom, it is enough to operate special anticanonical
transformations in the sector of original fields only.
In the case under consideration, we choose the generating function
of anticanonical transformations which acts in the sector of fields
$\varphi^{\mu\nu\lambda}$
and depends on the scalar field, $h^{\mu\nu\lambda}=h^{\mu\nu\lambda}(\phi)$.
Then, the most general form of
the generating function $h^{\mu\nu\lambda}$
with three derivatives responsible for the generation of quintic vertices
reads
\beq
\nonumber
&&h^{\mu\nu\lambda}=a_2\frac{1}{\Box}
\big[c_1\pa^{\mu}\pa^{\nu}\pa^{\lambda}\phi \;\phi^3+
c_2\pa^{(\mu}\pa^{\nu}\phi\;\pa^{\lambda)}\phi\;\phi^2+
c_3\pa^{\mu}\phi\;\pa^{\nu}\phi\;\pa^{\lambda}\phi\;\phi+
c_4\eta^{(\mu\nu}\pa^{\lambda)}\Box\phi\;\phi^3
+\\
\label{b3}
&&\qquad\qquad\quad+
c_5\eta^{(\mu\nu}\pa^{\lambda)}\pa_{\sigma}\phi\;\pa^{\sigma}\phi\;\phi^2+
c_6\eta^{(\mu\nu}\pa^{\lambda)}\phi\;\pa_{\sigma}\phi\;\pa^{\sigma}\phi\;\phi+
c_7\Box\phi\;\eta^{(\mu\nu}\pa^{\lambda)}\phi\;\phi^2\big],
\eeq
where $a_2$ is a coupling constant with ${\rm dim}(a_2)=-(3d-4)/2$ and
$c_i,\;\; i=1,2,...,7$
are arbitrary dimensionless constants.

The deformed action, $\widetilde{S}[\varphi,\phi]=S_0[\varphi+h,\phi]$,
can be presented in the form
\beq
\label{b4}
\widetilde{S}[\varphi,\phi]=S_0[\varphi,\phi]+S_{3\;loc}[\varphi,\phi] +
(non-local\;\; terms),
\eeq
where $S_{3\;loc}[\varphi,\phi]$ is the local functional
\beq
\nonumber
&&S_{3\;loc}[\varphi,\phi]=2a_2\int dx \varphi_{\mu\nu\lambda}
\Big[c_1\pa^{\mu}\pa^{\nu}\pa^{\lambda}\phi \;\phi^3+
c_2\pa^{(\mu}\pa^{\nu}\phi\;\pa^{\lambda)}\phi\;\phi^2+
c_3\pa^{\mu}\phi\;\pa^{\nu}\phi\;\pa^{\lambda}\phi\;\phi-\\
\nonumber
&&\qquad\qquad
-(c_1+c_4(d+1))\eta^{(\mu\nu}\pa^{\lambda)}\Box\phi\;\phi^3
-(2c_2+c_5(d+1))\eta^{(\mu\nu}\pa^{\lambda)}\pa_{\sigma}\phi\;\pa^{\sigma}\phi\;\phi^2-\\
&&\qquad\qquad
-(c_3+c_6(d+1))\eta^{(\mu\nu}\pa^{\lambda)}\phi\;
\pa_{\sigma}\phi\;\pa^{\sigma}\phi\;\phi-(c_2+c_7(d+1))
\Box\phi\;\eta^{(\mu\nu}\pa^{\lambda)}\phi\;\phi^2\Big],
\label{b5}
\eeq
describing interactions among the massless spin 3 and real scalar fields in the fifth
order.
By the same reasons explained in \cite{L,L1}, the gauge algebra
does not deform under special anticanonical transformations
(\ref{b3}) so that the deformed action is invariant under gauge transformations
(\ref{b2}), $\delta\widetilde{S}[\varphi,\phi]=0$. Due to the locality
of original gauge transformations, the action $S_{3\;loc}[\varphi,\phi]$ should be also
gauge-invariant, $\delta S_{3\;loc}[\varphi,\phi]=0$. Analysis of this requirement leads
to the following conditions on the constants $c_i,\;  i=1,...,7$,
\beq
\label{b6}
2c_2=c_3=6c_1,\quad c_4=-\frac{1}{2(d+1)}c_1,\quad
c_5=c_6=2c_7=-\frac{3}{(d+1)}c_1.
\eeq
We state that the functional
\beq
\nonumber
&&S_{3\;loc}[\varphi,\phi]=a_2\!\!\int \!\!dx \varphi_{\mu\nu\lambda}
\Big[\pa^{\mu}\pa^{\nu}\pa^{\lambda}\phi \;\phi^3+
3\pa^{(\mu}\pa^{\nu}\phi\;\pa^{\lambda)}\phi\;\phi^2+
6\pa^{\mu}\phi\;\pa^{\nu}\phi\;\pa^{\lambda}\phi\;\phi
-\frac{1}{2}\eta^{(\mu\nu}\pa^{\lambda)}\Box\phi\;\phi^3\\
&&\qquad\qquad\qquad
-3\eta^{(\mu\nu}\pa^{\lambda)}\pa_{\sigma}\phi\;\pa^{\sigma}\phi\;\phi^2
-3\eta^{(\mu\nu}\pa^{\lambda)}\phi\;
\pa_{\sigma}\phi\;\pa^{\sigma}\phi\;\phi-\frac{5}{2}
\Box\phi\;\eta^{(\mu\nu}\pa^{\lambda)}\phi\;\phi^2\Big],
\label{b7}
\eeq
is gauge-invariant under original gauge transformations (\ref{b2}),
\beq
\label{b8}
\delta S_{3\;loc}[\varphi,\phi]=0.
\eeq
and describes quintic vertices. The action $S_{3\;loc}[\varphi,\phi]$
is local and presents the exact form of quintic vertices  which cannot be
constructed using the standard Noether procedure adopted in the theory
of higher spin fields (for detailed description of relations between the standard Noether
procedure and new method see \cite{L}).

\section{Vertices $\sim\varphi A \phi\phi\phi$}
Here, we want to elaborate interactions among massless spin 3,
massive vector and real scalar fields in the framework of
deformation procedure \cite{BL-1,BL-2,L-2022} beginning with the
free initial action,
\beq
\label{c1}
S_0[\varphi,A,\phi]=S_0[\varphi,\phi]+S_0[A],
\eeq
where the action $S_0[\varphi,\phi]$ was defined in (\ref{b1}) and
\beq
\label{c2}
S_0[A]=-\int dx \Big(\frac{1}{4}F_{\mu\nu}F^{\mu\nu}+
\frac{1}{2}m_0^2A_{\mu}A^{\mu}\Big),\quad
F_{\mu\nu}=\pa_{\mu}A_{\nu}-\pa_{\nu}A_{\mu}
\eeq
is the action of massive vector field. The action (\ref{c1}) is invariant under
the gauge transformations
\beq
\label{c3}
\delta\varphi^{\mu\nu\lambda}=\pa^{(\mu}\xi^{\nu\lambda)},\quad
\eta_{\mu\nu}\xi^{\mu\nu}=0, \quad \delta A_{\mu}=0,\quad\delta\phi=0.
\eeq

The cubic ($\sim\varphi A \phi$)  and quartic ($\sim\varphi A \phi\phi$) vertices
for the initial gauge model (\ref{c1}) have been studied in \cite{L1}.
It was shown that the local cubic vertices invariant under original gauge transformations
are forbidden while the local quartic vertices are constructed in
the form of explicit functional which is exactly invariant under the initial gauge symmetry.

Consider now quintic vertices ($\sim\varphi A \phi\phi\phi$) in the model under
consideration. It is enough to chose the generating function of anticanonical transformations
in the form
\beq
\label{c4}
h^{\mu\nu\lambda}=a_2\frac{1}{\Box}K^{\mu\nu\lambda},
\eeq
where $K^{\mu\nu\lambda}=K^{\mu\nu\lambda}(A,\phi)$ is the local function
\beq
\nonumber
&&K^{\mu\nu\lambda}=
c_1\pa^{(\mu}\pa^{\nu}A^{\lambda)}\;\phi^3+
c_2\pa^{(\mu}A^{\nu}\;\pa^{\lambda)}\phi\;\phi^2+
c_3A^{(\mu}\;\pa^{\nu}\pa^{\lambda)}\phi\;\phi^2
+c_4A^{(\mu}\;\pa^{\nu}\phi\;\pa^{\lambda)}\phi\;\phi+\\
\nonumber
&&\qquad\qquad+
c_5\eta^{(\mu\nu}\Box A^{\lambda)}\;\phi^3
+c_6\eta^{(\mu\nu}\pa_{\sigma}A^{\lambda)}\;\pa^{\sigma}\phi\;\phi^2+
c_7\Box\phi\;\eta^{(\mu\nu}A^{\lambda)}\;\phi^2+
c_8\eta^{(\mu\nu}A^{\lambda)}\;\pa_{\sigma}\phi\;\pa^{\sigma}\phi\;\phi+\\
\nonumber
&&\qquad\qquad+c_9\eta^{(\mu\nu}\pa_{\lambda)}\pa_{\sigma}A^{\sigma}\;\phi^3+
c_{10}\eta^{(\mu\nu}\pa^{\lambda)}\phi\;\pa_{\sigma}A^{\sigma}\;\phi^2+
c_{11}\eta^{(\mu\nu}\pa^{\lambda)}A^{\sigma}\;\pa_{\sigma}\phi\;\phi^2+\\
&&\qquad\qquad+c_{12}\eta^{(\mu\nu}\pa^{\lambda)}\pa_{\sigma}\phi\;A^{\sigma}\;\phi^2+
c_{13}\eta^{(\mu\nu}\pa^{\lambda)}\phi\;A^{\sigma}\;\pa_{\sigma}\phi\;\phi ,
\label{c5}
\eeq
$a_2$ is a coupling constant with ${\rm dim}(a_2)=-(3d-6)/2$ and $c_i,\; i=1,2,...,13$
are dimensionless constants. Modification of local part of the deformed theory in terms of
$K^{\mu\nu\lambda}$ is described by the functional
\beq
\label{c6}
S_{3\; loc}[\varphi,A,\phi]=2a_2\int dx \varphi_{\mu\nu\lambda}\big[K^{\mu\nu\lambda}-
\eta^{(\mu\nu}K^{\lambda)\rho\sigma}\eta_{\rho\sigma}\big].
\eeq
Note that the gauge algebra does not deform under anticanonical transformations generated by
the function $h^{\mu\nu\lambda}$ (\ref{c4}), (\ref{c5}). It leads to the requirement for the
functional (\ref{c6}) to be invariant under original gauge transformations (\ref{c3}),
\beq
\label{c7}
\delta S_{3\; loc}[\varphi,A,\phi]=0.
\eeq
Analysis of this equation gives the following conditions on constants $c_i,\; i=1,2,...,13$
\beq
\nonumber
&&c_2=2c_3=c_4=6c_1,\quad c_5=c_6=c_7=c_8=0,\\
&&c_9=3c_{10}=3c_{11}=3c_{12}=
6c_{13}=-\frac{3}{2(d+1)}c_1.
\label{c8}
\eeq
As a result,  the local functional
\beq
\nonumber
&&S_{3\;loc}[\varphi,A,\phi]=2a_2\!\int \!dx \varphi_{\mu\nu\lambda}
\Big[\pa^{(\mu}\pa^{\nu}A^{\lambda)}\;\phi^3+
6\pa^{(\mu}A^{\nu}\;\pa^{\lambda)}\phi\;\phi^2+
3A^{(\mu}\;\pa^{\nu}\pa^{\lambda)}\phi\;\phi^2+\\
\nonumber
&&\qquad\qquad\quad
+6A^{(\mu}\;\pa^{\nu}\phi\;\pa^{\lambda)}\phi\;\phi
-\eta^{(\mu\nu}\Box A^{\lambda)}\;\phi^3
-6\eta^{(\mu\nu}\pa_{\sigma}A^{\lambda)}\;\pa^{\sigma}\phi\;\phi^2-
3\Box\phi\;\eta^{(\mu\nu}A^{\lambda)}\;\phi^2-\\
\nonumber
&&\qquad\qquad\quad
-6\eta^{(\mu\nu}A^{\lambda)}\;\pa_{\sigma}\phi\;\pa^{\sigma}\phi\;\phi-
-\frac{1}{2}\eta^{(\mu\nu}\pa^{\lambda)}\pa_{\sigma}A^{\sigma}\;\phi^3-
\frac{11}{2}\eta^{(\mu\nu}\pa^{\lambda)}\phi\;\pa_{\sigma}A^{\sigma}\;\phi^2-\\
&&\qquad\qquad\quad-
\frac{11}{2}\eta^{(\mu\nu}\pa^{\lambda)}A^{\sigma}\;\pa_{\sigma}\phi\;\phi^2-
\frac{5}{2}\eta^{(\mu\nu}\pa^{\lambda)}
\pa_{\sigma}\phi\;A^{\sigma}\;\phi^2-
\frac{47}{4}\eta^{(\mu\nu}\pa^{\lambda)}\phi\;A^{\sigma}\;\pa_{\sigma}\phi\;\phi\Big]
\label{c9}
\eeq
describes quintic vertices.
These vertices are invariant under original gauge transformations and may be considered as
the first explicit construction of interactions of the fifth order in the theory of higher
spin fields.

\section{Discussion}
In the present paper, quintic vertices  describing interactions
among massless spin 3 field and massive vector and real scalar fields
have been constructed. The construction was based on the new approach to the deformation
procedure \cite{BL-1,BL-2,L-2022}.

Main advantage of the new approach is related with possibility
to present the deformation of a given dynamical system with gauge freedom
in a closed and explicit form using a generating function of anticanonical transformations
of the BV formalism. Final results of the deformation are formulated in a simple enough way.
Namely, if $S_0[A]$ is an initial action of fields $A^i$ invariant under gauge transformations
$\delta A^i=R^i_{\alpha}(A)\xi^{\alpha}$ then the deformed action $\widetilde{S}[A]$ is
constructed by the rule $\widetilde{S}[A]=S_0[A+h(A)]$, where $h^i(A)$ is the generating
function of anticanonical transformations which act non-trivially in the space of initial
fields $A^i$ only. The deformed action is invariant under deformed gauge transformations
$\widetilde{\delta}A^i=(M^{-1}(A))^i_{\;j}R^j_{\alpha}(A+h(A))\xi^{\alpha}$,
$\widetilde{\delta}\widetilde{S}[A]=0$, where $(M^{-1}(A))^i_{\;j}$ is the matrix inverse to
$(M(A))^i_{\;j}=\delta^i_{\;j}+h^i(A)\overleftarrow{\pa}_{\!\!A^j}$.

The vertices considered in this paper were constructed using very special anticanonical
transformations that do not deform the gauge algebra.
In turn, this made it possible to quite simply single out the local part
of the deformed action
and analyze the conditions for its invariance.
Interest to results obtained in the paper may be connected at least with two reasons.
Firstly, in our knowledge, quintic vertices  did not construct for any
models of the theory of higher
spin fields. Secondly, in connection with the  problem of locality in the high
spin theory \cite{JT,Vas,Did}, examples (\ref{b7}), (\ref{c9}) give
 local gauge-invariant vertices, which allow to assume the existence of local vertices
in higher orders of the perturbation theory.

\section*{Acknowledgments}
\noindent
We thank I.L. Buchbinder and M.A. Vasiliev for useful discussions.
The work of PML is supported by the Ministry of
Education of the Russian Federation, project FEWF-2020-0003.

\end{document}